\documentclass[iop]{emulateapj}
\usepackage{natbib}
\usepackage{bm}
\usepackage{gensymb}
\usepackage[usenames]{color}
\definecolor{violet}{rgb}{0.6,0.0,0.3}
\usepackage[%
pdftitle={Probing Dark Energy with the Kunlun Dark Universe
Survey Telescope},%
pdfauthor={Zhao et al.},%
bookmarksnumbered=true,%
linkcolor=violet,citecolor=blue,urlcolor=blue,colorlinks=true]{hyperref}

\newcommand{\be}{\begin{equation}}
\newcommand{\ee}{\end{equation}}
\newcommand{\ba}{\begin{eqnarray}}
\newcommand{\ea}{\end{eqnarray}}

\newcommand{\bit}[1]{\mbox{\textbf{\emph{#1}}}}

\newcommand{\phz}{photo-\emph{z}}

\newcommand{\remove}[1]{}

\begin{document}

\title{Probing Dark Energy with the Kunlun Dark Universe
Survey Telescope}

\author{Gong-Bo Zhao$^{1,2}$}
\author{Hu Zhan$^3$}
\author{Lifan Wang$^{4,5}$}
\author{Zuhui Fan$^6$}
\author{Xinmin Zhang$^{1,7}$}

\shortauthors{Zhao et al.} \shorttitle{Probing Dark Energy with
KDUST}

\affiliation{ $^1$~Theoretical Physics Division,
Institute of High Energy Physics, Chinese Academy of Sciences\\
P.O.Box 918-4, Beijing 100049, P.~R.~China\\
 $^2$~Institute of Cosmology and
Gravitation, University of Portsmouth, Dennis Sciama Building\\ Burnaby Road, Portsmouth, PO1
3FX, United Kingdom \\
$^3$~Key Laboratory for Optical Astronomy, National Astronomical
Observatories, \\
Chinese Academy of Sciences, Beijing 100012, P.~R.~China\\
$^4$~Physics Department, Texas A\&M University, College Station, TX
77843, USA\\
$^5$~Purple Mountain Observatory, Nanjing 210008, P.~R.~China\\
$^6$~Department of Astronomy, School of Physics, Peking University,
Beijing 100871, P.~R.~China \\
$^7$~Theoretical Physics Center for Science Facilities,
Chinese Academy of Sciences, P.~R.~China }

\begin{abstract}
Dark energy is an important science driver of many upcoming
large-scale surveys. With small, stable seeing and low thermal infrared
background, Dome A, Antarctica, offers a unique opportunity for
shedding light on fundamental questions about the universe. We show
that a deep, high-resolution imaging survey of 10,000 square
degrees in \emph{ugrizyJH} bands can provide  competitive
constraints on dark energy equation of state parameters using type
Ia supernovae, baryon acoustic oscillations, and weak lensing
techniques. Such a survey may be partially achieved with a coordinated
effort of the Kunlun Dark Universe Survey Telescope (KDUST)
in \emph{yJH} bands over 5000--10,000 deg$^2$ and
the Large Synoptic Survey Telescope in
\emph{ugrizy} bands over the same area. Moreover,
the joint survey can take advantage of
the high-resolution imaging at Dome A to further tighten the constraints on
dark energy and to measure dark matter properties with strong lensing
as well as galaxy--galaxy weak lensing.
\end{abstract}


\section{Introduction}

Antarctic Plateau, especially the Kunlun Station under construction
by the Polar Research Institute of China (PRIC), provides a unique
opportunity for wide field astronomical surveys targeting
cosmological studies. Astronomical site survey of Dome A, Antarctica
was enabled by the International Polar Year (IPY) endorsed PANDA
program~\citep{Yang09} led by the PRIC and the Chinese Center
for Antarctica Astronomy (CCAA). Several international teams
contributed to this effort. In particular, the power and on-site
laboratory system built by the University of New South Wales (BUN'S)
has provided the platform for all the site survey instruments.

In its two years' operation, the site
survey effort proves practically all aspects of the theoretical
expectations of the Dome A site for astronomical observations.
Preliminary analyses show that the boundary layer of atmospheric
turbulence to be around $10-20$ meters during the Antarctic winter~\citep{ashley10}. Similar to the relatively better
studied neighboring Dome C site~\citep{fossat10}, the Dome A site
may enjoy free atmospheric seeing conditions of about 0.3 arcsec
seeing above this boundary layer, thus making it an ideal site for
high angular resolution wide area surveys.

The temperature at Dome A
is around $-60$ to $-70 \degree{C}$, making it the coldest spot on
the surface of the Earth. This implies a very low thermal background
emission in the thermal infrared.
The Dome A site is thus also the best site for astronomical
observations in the near infrared wavelength region.

One other
especially exciting property of the site is the lack of water vapors
due to the high altitude and the low temperature. This implies that
the site is also ideal for terahertz observations which have been
impossible from any temperate sites on Earth.

While quantitative
site properties are still under analyses and longer term monitoring
are still needed to firmly establish the astronomical potential of
the Dome A site, there is no doubt that a survey project
at Dome A can be highly complimentary to programs such as the Large
Synoptic Survey Telescope\footnote{See \url{http://www.lsst.org/}.}
\citep[LSST,][]{lsst09}, the Joint Dark Energy
Mission\footnote{See \url{http://jdem.gsfc.nasa.gov/}.}, and
Euclid\footnote{See \url{http://sci.esa.int/euclid/}.}.
For example, a survey at Dome A can provide
near infrared data that compliments the deep optical band survey of
the LSST; a deep survey in the near infrared combined with the
optical data from LSST can reveal high redshift objects at $z\sim10$,
which are not detectable in the optical. The Kunlun Dark
Universe Survey Telescope (KDUST) is a 6-to-8-meter wide-area survey
telescope being designed by the CCAA. The preliminary design
includes a $3\times3$ square degree optical camera with $0''.15$
pixel, and an infrared camera of $1\times1$ square degree at
$0''.1$/pixel optimized for $1-3.5$ $\mu$m surveys.

One of the key science missions of KDUST is to investigate the
mystery of the accelerated cosmic expansion
\citep{riess98,perlmutter99a} using multiple techniques. In this
paper, we estimate how well an ideal 10,000 deg$^2$
\emph{ugrizyJH} survey can constrain the dark energy equation of
state (EOS) with weak lensing (WL), baryon acoustic oscillations
(BAOs), and type Ia supernova (SNe) luminosity distances. These dark
energy probes have different sensitivities to the cosmic expansion
and structure growth as well as various systematic uncertainties in
the observations, and hence are highly complementary to each other
for constraining dark energy properties
\citep[e.g.,][]{knox05,zhan06d,zhan09}.

The SNe technique relies on the standardizable candle of the SNe
intrinsic luminosity \citep{phillips93} to measure the luminosity
distance, $D_{\rm L}(z)$. Dark energy properties can then be
inferred from the distance--redshift relation. The BAO technique
utilizes the standard ruler of the baryon imprint on the matter (and
hence galaxy) power spectrum \citep{peebles70, bond84} to measure
the angular diameter distance, $D_{\rm A}(z)$, and, if the redshifts
are sufficiently accurate, the Hubble parameter, $H(z)$
\citep{eisenstein98, cooray01b, blake03, hu03b, linder03, seo03}.
The WL technique has the
advantage that it can measure both $D_{\rm A}(z)$ from the lensing
kernel and the growth factor of the large-scale structure $G(z)$
\citep{hu99,huterer02b,refregier03,takada04,knox06b,zhan09}.

Because of the excellent seeing condition and infrared accessibility
at Dome A, KDUST has a number of advantages for the commonly used
cosmological probes. For example, the signal-to-noise ratio for
point sources is inversely proportional to the seeing.
Thus, a 6 meter telescope at Dome A ($\sim 0''.3$ median seeing
in the optical) would be
equivalent to a 14 meter telescope at a temperate site ($\sim 0''.7$
seeing) for point-source observations with the same sky background
level. An 8m KDUST could detect SNe out to redshift 3 in the
$K_{\rm dark}$ ($2.27$--$2.45\mu$m, redward of $K$) band
\citep{kim2010}. Although high-z distances are not sensitive to
conventional dark energy, they can be used to determine the mean
curvature accurately, which, in turn, helps constrain dark energy
EOS \citep{linder05b,knox06c}. Moreover, even though dark energy is
thought to be sub-dominant at high redshift, there is no direct
evidence to prove one way or another. Measurements of SNe at $z > 2$
will provide crucial data for tests of early dark energy.

Small and stable seeing is particularly helpful for WL. One
could resolve more galaxies at the same surface brightness limit,
which reduces the shape noise for shear measurements. Fine
resolution helps measure the shape accurately and reduce the shear
measurement systematic errors. In addition, deep \emph{JH}
photometry can track the 4000\AA{} break of an elliptical galaxy
to $z \sim 3$ and
improve photometric redshifts (\phz{}s) as well as systematic
uncertainties in the \phz{} error distribution \citep{abdalla08},
which has a large impact on
WL constraints on the dark energy EOS \citep{huterer06,ma06,zhan06d}.

Adding \emph{K} or $K_{\rm dark}$ band will certainly improve galaxy
\phz{}s, especially at $z \gtrsim 3$. However, currently planned
multiband dark energy surveys use galaxies at $z \lesssim 3$, and their
concern is the confusion between $z\lesssim 0.5$ ellipticals and
$2 \lesssim z \lesssim 3.5$ star-forming galaxies, which is greatly
mitigated by $u$ and $JH$ bands \citep{abdalla08}.
Another consideration is that Dome A is far more advantageous at
$K_{\rm dark}$ band than at $K$ band because of the low thermal
background there, so $K_{\rm dark}$ is likely to be chosen over $K$.
This would leave a considerable gap in the wavelength coverage and
reduce the already-small gain on \phz{}s in the useful redshift
range for dark
energy investigations. Therefore, we do not discuss utilities of
wavebands beyond $H$ in this paper.
Nevertheless, the $K_{\rm dark}$ band is crucial
for a broad range of other sciences and will be an important aspect
of the KDUST survey.

This paper is organized as follows. Section~\ref{sec:survey}
discusses the survey plan of KDUST including that of its pathfinder
in context of the LSST survey. We then consider a joint KDUST
and LSST survey in \autoref{sec:de} for constraining
the dark energy EOS with BAO, WL, and type Ia SN techniques. The
results are presented in both two-parameter space where
the dark energy EOS is parameterized as $w(z)=w_0+w_az(1+z)^{-1}$
and in model-independent principle component space. The conclusion is
drawn in \autoref{sec:con}.

\section{A Potential KDUST Survey Plan} \label{sec:survey}

Dome A has a great potential for cosmology, but, given other ambitious
projects that will be concurrent with KDUST, one must carefully plan
the KDUST survey to make the best use of the Dome A site. As we discuss,
below, a potentially efficient strategy for KDUST is to focus on the
near infrared (NIR)
bands and incorporate optical data from LSST or other surveys.

In the optical bands, a 6m KDUST is about twice as fast as LSST for
surveying sky-dominated point sources at the same sky level, in which
case the survey speed is proportional to the aperture and field of view
and inversely proportional to the seeing disk area.
In reality, aurorae increase the sky brightness
in short wavelengths. It is estimated that the sky brightness at
Dome A would be twice as bright as that at the best temperate site
in $B$ and 20\%--30\% brighter in $V$ \citep{saunders09};
measurements have  shown that the median $i$-band sky brightness
at Dome A in 2008\footnote{The sky background was seriously affected by
the Moon in 2008, as it was always close to full when above the horizon.}
was $19.81$ mag arcsec$^{-2}$ and $20.46$ mag arcsec$^{-2}$ during
dark time, better than that at other good sites \citep{zou10}.
With the above considerations, the
6m KDUST could survey 10,000 deg$^2$ to LSST depths in $griz$ and much
deeper in $y$ ($\sim 26$ mag, $5\sigma$ point sources) in 2.5 years.
The NIR camera of KDUST would have a
much smaller field of view. It could reach $J = 25$ mag and
$H = 24.6$ mag over the same area in 3.5 years.

Since LSST plans to survey the southern sky in $ugrizy$, it is not
absolutely necessary for KDUST to survey in the optical again except
in the $y$ band. LSST would spend 20\% of its time in $y$ band to
achieve a 5-$\sigma$ limiting magnitude of 24.4 for point sources,
which is 2.8 magnitudes shallower than its $r$ band limit. From \phz{}
consideration, it is desirable to have the $y$ band limit not too
much shallower than the limits in shorter wavebands. KDUST could
improve the situation in its 10,000 deg$^2$ survey area, which would
be covered by LSST as well.
A joint effort of KDUST in $yJH$ and LSST in
$ugrizy$ would save both projects time while achieving better
performance in measuring the dark energy EOS.

Here, we use SNAP and LSST as references to model an ideal 10,000
square degree survey (named ``ideal 10k'') in \emph{ugrizyJH} bands
to comparable depth as SNAP. We assume that its galaxy
redshift distribution follows
\[
n(z) \propto z^2 e^{-z/z_*}
\]
with $z_* = 0.5$. The projected galaxy number density $\bar{n}_g$
is 70 arcmin$^{-2}$, and the distribution peaks at $2z_*$.
For LSST, we adopt $z_* = 0.3$ and $\bar{n}_g = 40$ arcmin$^{-2}$
\citep{lsst09}. The model parameters of the survey data are
summarized in \autoref{tab:suv}.

The galaxy distribution $n_i(z)$ in the $i$th bin
is sampled from $n(z)$ by \citep{ma06,zhan06d}
\[
n_i(z) = n(z) \mathcal{P}(z_{{p},i}^{\rm B}, z_{{p},i}^{\rm E}; z),
\]
where the subscript p denotes \phz{} space,
$z_{{p},i}^{\rm B}$ and $z_{{p},i}^{\rm E}$ define the
extent of bin $i$, and $\mathcal{P}(a,b;z)$ is the probability of
assigning a galaxy that is at true redshift $z$ to
the \phz{} bin between $z_{p} = a$ and $b$.
We approximate the \phz{} error to be Gaussian with bias $\delta z$
and rms $\sigma_z=\sigma_{z0} (1+z)$, and the probability becomes
\begin{eqnarray} \nonumber
\mathcal{P}(z_{{p},i}^{\rm B}, z_{{p},i}^{\rm E}; z)
&=& I(z_{{p},i}^{\rm B}, z_{{p},i}^{\rm E}; z)
 / I(0,\infty; z), \\ \nonumber
I(a, b; z) &=& \frac{1}{\sqrt{2\pi}\,\sigma_z} \int_a^b d z_{p}
\,\exp\left[-\frac{(z_{p} - z - \delta z)^2}{2\sigma_z^2}\right].
\end{eqnarray}
The normalization $I(0,\infty; z)$ implies that galaxies with
a negative \phz{} have been excluded from $n(z)$. We use 40 \phz{}
bias $\delta z$ and 40 \phz{} rms $\sigma_z$ parameters evenly
spaced between $z = 0$ and 5 to model the \phz{} error
distribution in the $z$--$z_p$ space; the \phz{} bias and rms at
any redshift are linearly interpolated from these 80 parameters.
Note that the \phz{} parameters are assigned independent of galaxy
bins. We assume that $\sigma_{z0} = 0.03$ for the ideal survey
and $\sigma_{z0} = 0.05$ for LSST.

\begin{table}
\centering 
\caption{Model parameters of survey data and priors. \label{tab:suv}}
\begin{tabular}{c|c|c|cc}
 \hline  \hline
&  Ideal 10k & LSST & \multicolumn{2}{c}{KDUST+LSST\tablenotemark{a}} \\
 \hline
area (deg$^2$) & 10,000 & 20,000 & \multicolumn{2}{c}{10,000\,+\,10,000} \\
$n_{\rm g}$ (arcmin$^{-2}$) & 70 & 40 & 60 & 40\\
$z_*$     & 0.5 & 0.3 & 0.4 & 0.3 \\
$\sigma_{z0}$  & 0.03 & 0.05 & 0.03 & 0.04\\
$\sigma_{\rm P}(\delta z)$ & $0.2 \sigma_z$ & $0.3\sigma_z$ &
  $0.2\sigma_z$ & $0.2\sigma_z$ \\
$\sigma_{\rm P}(f^\gamma_i)$ & 0.002 & 0.005 & 0.002 & 0.003 \\
$(A_i^\gamma)^2 (\times10^{-10})$ & $4$ & $10$ & $4$ & $6$ \\
 \hline
 \hline
\end{tabular}
\tablenotetext{1}{We assume that KDUST would survey 10,000 deg$^2$
in \emph{yJH} bands and several hundred deg$^2$ in \emph{griz} bands
around the south celestial pole, excluding areas with heavy galactic
extinction. In the 10,000 deg$^2$ overlap region
between KDUST and LSST, the combined survey is close to the
ideal 10k survey. The data outside the overlap region come solely
from LSST but are expected to have smaller systematics than they do
without KDUST because of better calibrations from the overlap region.}
\vspace{1ex}
\end{table}

Uncertainties of the \phz{} parameters (or, the \phz{} error
distribution in general) have a large impact on
the dark energy constraints from WL
and are referred to as \phz{} systematics. Therefore, the prior on
the \phz{} error distribution is an important quantity to specify
when reporting WL constraints on dark energy. To reduce the
dimension of the investigation, we set a simple function for all
the priors on the \phz{} bias parameters,
$\sigma_\mathrm{P}(\delta z) = 0.2 \sigma_z$, and peg the priors
on \phz{} rms parameters to those on the bias parameters:
$\sigma_\mathrm{P}(\sigma_z) = \sqrt{2} \sigma_\mathrm{P}(\delta z)$.
For Gaussian \phz{} errors, these priors correspond to a calibration
sample of 25 spectra per redshift interval of the \phz{} parameters.
We set $\sigma_\mathrm{P}(\delta z) = 0.3 \sigma_z$, reflecting
larger uncertainties with fewer filter bands. However, such difference
in the \phz{} priors is not important if one performs a joint analysis
of the BAO and WL techniques, which takes advantage of the
self-calibration of \phz{} error distribution by galaxy power
spectra \citep[referenced herein]{schneider06, zhan06d, zhang09}

LSST will benefit from KDUST $yJH$ data in a number of ways. In the 10,000
deg$^2$ overlap region, KDUST data will (1) improve \phz{}s directly,
(2) increase the galaxy sample with those that would not meet
the LSST optical photometry selection criteria, or whose \phz{} could
not be reliably determined in the absence of the deep $yJH$ data,
or whose shape
could not be measured well because of the larger seeing at the LSST
site, and (3) provide calibration of various systematic errors such as
those in shear measurements. Even in the non-overlap region, LSST
could still improve \phz{}s and shear measurements because of the
calibration within the overlap region. This is reflected in the last
column of \autoref{tab:suv} where we construct a joint survey combining
LSST and KDUST.

For the SN data, we assume that KDUST would obtain a SNAP-like sample
of 2000 SNe reaching redshift $1.7$ as well as a sample of 1000 local
and nearby SNe. Details of the assumptions for
the SNAP experiments and
the calculation of the Fisher matrices can be found in
\citep{Pogosian:2005ez} and in \citep{Zhao:2008bn}.

\begin{figure*}[htp]
\centering
  \includegraphics[width=0.7\textwidth]{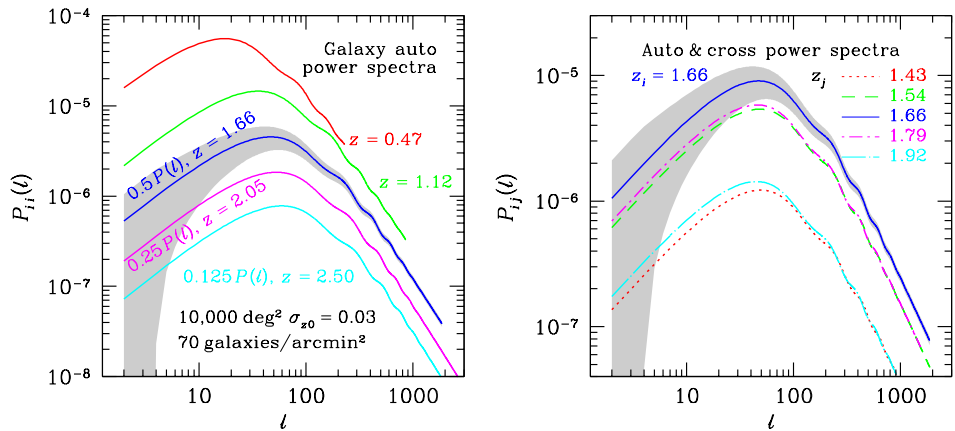}
  \caption{\emph{Left panel}: Five galaxy auto power spectra from the
ideal 10k survey. The BAO feature is visible around $\ell\sim 400$ in
the highest redshift bin and move toward lower multipoles in lower
redshift bins. Note, however, that the linear scale of the BAOs is
fixed in the comoving frame. The amplitude of the power spectra
and that of the BAO feature decreases as the \phz{} rms error
increases. The shaded area gives the 1-$\sigma$
error (e.g., sample variance and shot noise) of the power spectrum
\emph{per} $\ell$ in the bin centered at $z = 1.66$. The power spectra
are shifted for clarity.
\emph{Right panel}: Cross power spectra between the bin at $z = 1.66$
and its second neighboring bins and 4th neighboring bins.
 \label{fig:bcl}}
\end{figure*}

\begin{figure*}[htp]
\centering
  \includegraphics[width=0.7\textwidth]{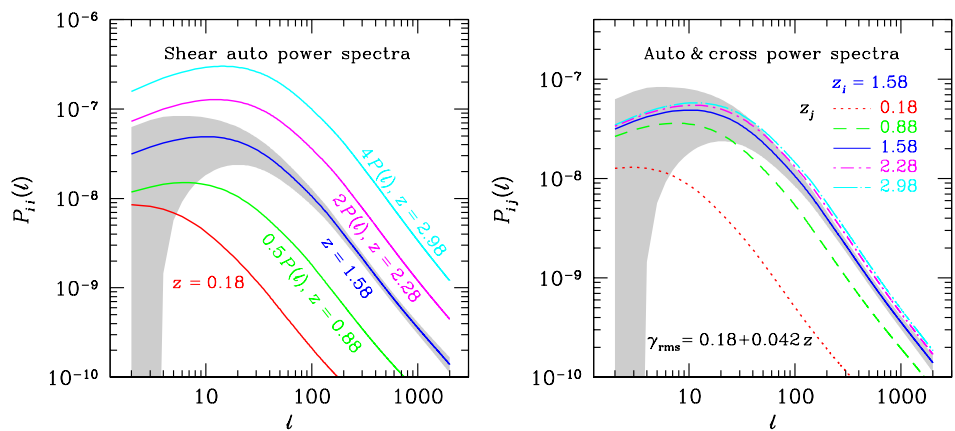}
  \caption{Same as \autoref{fig:bcl}
but for the shear power spectra in 5 different redshift bins. The
shear power spectra are smoother than the galaxy power spectra,
because lensing kernels are much broader that galaxy distributions
in the \phz{} bins.
 \label{fig:wcl}}
\end{figure*}

\section{Constraining Dark Energy with KDUST and LSST} \label{sec:de}

In this section, we estimate the constraints on the dark energy EOS
from the joint survey of KDUST and LSST using BAO, WL, and SN
techniques.

\subsection{Baryon Acoustic Oscillations and Weak Lensing}

The angular power spectra of
the galaxy number density $n(\boldsymbol{\theta})$ and
(E-mode) shear $\gamma(\boldsymbol{\theta})$ can be written as
\citep{hu04b,zhan06d}
\be \label{eq:aps}
P_{ij}^{XY}(\ell) = \frac{2\pi^2}{c\ell^3} \int d z\, H(z)
D_{\rm A}(z) W_i^X(z) W_j^Y(z) \Delta^2_\delta(k; z),
\ee
where lower case subscripts correspond to the tomographic bins, upper
case superscripts label the observables, e.g., $X= \mbox{g}$ for
galaxies or $\gamma$ for shear,
$\Delta^2_\delta(k;z)$ is the dimensionless power spectrum
of the density field, and $k = \ell/D_{\rm A}(z)$. The window
functions are
\ba 
W_i^{\rm g}(z) &=& b(z)\frac{n_i(z)}{\bar{n}_i}, \nonumber \\
W_i^\gamma(z) &=& \frac{3}{2} \frac{\Omega_{\rm
m}H_0^2}{H(z)}\frac{D_{\rm A}(z)}{a\,c} \int_z^\infty \! d z'\,
\frac{n_i(z')}{\bar{n}_i} \frac{D_{\rm A}(z,z')}{D_{\rm A}(z')},
\nonumber \ea where $b(z)$ is the linear galaxy clustering bias, and
$\Omega_{\rm m}$ and $H_0$ are, respectively, the matter fraction at
$z = 0$ and Hubble constant. The galaxy redshift distribution
$n_i(z)$ in the $i$th tomographic bin is an average of the
underlying three-dimensional galaxy distribution over angles, and
the mean surface density $\bar{n}_i$ is the total number of galaxies
per steradian in bin $i$. For WL, we use 10 bins evenly spaced
between $z = 0$ to 3.5, and for BAO, we use 30 bins from $z = 0.15$
to 3.5 with bin width proportional to $1+z$.

The $1\sigma$ statistical error of the mean power spectrum in the
multipole range $(\ell,\ell+\Delta\ell)$ is given by
\ba \nonumber
\Delta P_{ij}^{XY}(\ell) &=& \sqrt{\frac{2}{(2\ell+1) \Delta \ell
f_{\rm sky}}} \tilde{P}_{ij}^{XY}(\ell), \\ \nonumber
\tilde{P}_{ij}^{XY}(\ell) &=& P_{ij}^{XY}(\ell) + \delta_{XY}^{\rm
K}\delta_{ij}^{\rm K} \frac{X_{\rm rms}^2}{\bar{n}_i},
\ea
where
$\delta_{ab}^{\rm K}$ is the Kronecker delta function, $f_{\rm sky}$
is the sky coverage, ${\rm g_{\rm rms}}\equiv 1$, and $\gamma_{\rm
rms} = 0.18+0.042z$.

\autoref{fig:bcl} shows several examples of the galaxy auto and cross
power spectra, and \autoref{fig:wcl} shows shear auto and cross power
spectra. Cross power spectra between foreground galaxy bins and
background shear bins, $P_{ij}^{{\rm g}\gamma}$, are not shown, but
they have similar characteristic shapes of those in \autoref{fig:bcl}
and \autoref{fig:wcl}. The amplitude of the galaxy power spectra and
the strength of the BAO feature decreases with increasing \phz{} rms
error, because a larger \phz{} rms means more smoothing in the radial
direction. The amplitude of the shear power spectrum increases as
redshift increases, because higher redshift galaxies are lensed by
more intervening matter and, hence, have stronger shear signal
fluctuations. The amplitude of the galaxy cross power spectrum between
two bins is very sensitive to the separation between the two bins
in true-redshift space. Hence, galaxy cross power spectra can be used
to calibrate the \phz{} error distribution.

\subsection{Constraints on $w_0$ and $w_a$}

We use the Fisher information matrix \citep{tegmark97b} to estimate
the errors of the parameters of interest. In summary, the Fisher
matrix of the parameter set $\{q_\alpha\}$ is given by
\be \label{eq:trfish}
F_{\alpha\beta} = f_{\rm sky} \sum_\ell \frac{2\ell + 1}{2} {\rm Tr}\,
\bit{C}_\ell^{-1} \frac{\partial \bit{C}_\ell}{\partial q_\alpha}
\bit{C}_\ell^{-1} \frac{\partial \bit{C}_\ell}{\partial q_\beta},
\ee
with $(\bit{C}_\ell)_{ij}^{XY} = \tilde{P}_{ij}^{XY}(\ell)$
for galaxies and shear.
The minimum marginalized error of $q_\alpha$ is $\sigma(q_\alpha) =
(F^{-1})_{\alpha\alpha}^{1/2}$. Independent Fisher matrices are
additive; a prior on $q_\alpha$, $\sigma_{\rm P}(q_\alpha)$,
can be introduced via $F_{\alpha\alpha}^{\rm new} = F_{\alpha\alpha}
+\sigma_{\rm P}^{-2}(q_\alpha)$.

We extend the additive and multiplicative shear power spectrum errors
in \citet{huterer06} to include the galaxy power spectrum errors:
\begin{eqnarray} \nonumber
(\bit{C}_\ell^{XY})_{ij} &=& (1 + \delta_{X\gamma}^{\rm K} f_i^X +
\delta_{Y\gamma}^{\rm K} f_j^Y) P_{ij}^{XY}(\ell) + \\
&& \delta_{XY}^{\rm K} \left[\delta_{ij}^{\rm K}
\frac{X_{\rm rms}^2}{\bar{n}_i} + \rho^X A_i^X A_j^Y
\left(\frac{\ell}{\ell_*^X}\right)^{\eta^X}\right],
\end{eqnarray}
where $\rho^X$ determines how strongly the additive errors of two
different bins are correlated, and $\eta^X$ and $\ell_*^X$ account
for the scale dependence of the additive errors.
Note that the multiplicative error of galaxy number density is
degenerate with the galaxy clustering bias and is hence absorbed
by $b_i$. At the levels of systematics future surveys aim to
achieve, the most important aspect of the (shear) additive error is
its amplitude \citep{huterer06}, so we simply fix $\rho^X = 1$
and $\eta^X = 0$. For more comprehensive accounts of the above
systematic uncertainties, see
\citet{huterer06, jain06, ma06, zhan06d}.

Forecasts on dark energy constraints are sensitive to the priors
on the shear multiplicative errors ($\sigma_\mathrm{P}(f_i^\gamma)$)
and the amplitudes of the shear additive errors ($A_i^\gamma$),
so we list them in \autoref{tab:suv}. See
\citet{wittman05,massey07a,paulin-henriksson08} for detailed work
on these systematic uncertainties and \citet{zhan09} for a discussion
on $\sigma_\mathrm{P}(f_i^\gamma)$ and $A_i^\gamma$ for LSST. It has
been demonstrated that small and stable point spread function
in space leads to better
shear measurements \citep{kasliwal08}. In the absence of a detailed
investigation, we simply choose values or priors of these systematic
error parameters somewhat arbitrarily between what might be achieved
by space projects and what have been used in LSST forecasts.
We infer from the Sloan Digital
Sky Survey galaxy angular power spectrum \citep{tegmark02} that the
additive galaxy power spectrum error due to extinction, photometry
calibration, and seeing will be at $(A_i^\mathrm{g})^2 = 10^{-8}$ level.
Since the amplitude of $A_i^\mathrm{g}$ is fairly low
compared to the galaxy power spectra, its value does not affect the
forecasts much.

\begin{figure}
\centering
  \includegraphics[height=0.36\textwidth]{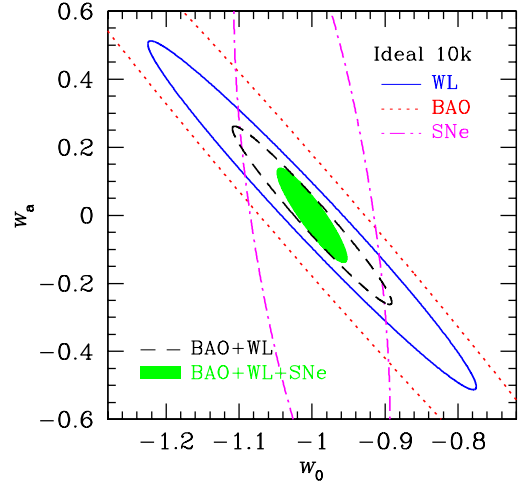}
  \caption{Forecasts of 1$\sigma$ errors on the dark energy
EOS parameters $w_0$ and $w_{\rm a}$ for the ideal 10k survey
WL (solid line), BAOs
(dotted line), SNe (dashed line), and the three combined (shaded
area). We have included \emph{Planck} priors in all the results.
Although the CMB priors have a significant impact on the SN results
and to a lesser degree on WL results and BAO results, they have a
much smaller effect on the joint constraints of BAO+WL and BAO+WL+SNe.
\label{fig:kdall}}
\end{figure}

\begin{figure}
\centering
\includegraphics[height=0.36\textwidth]{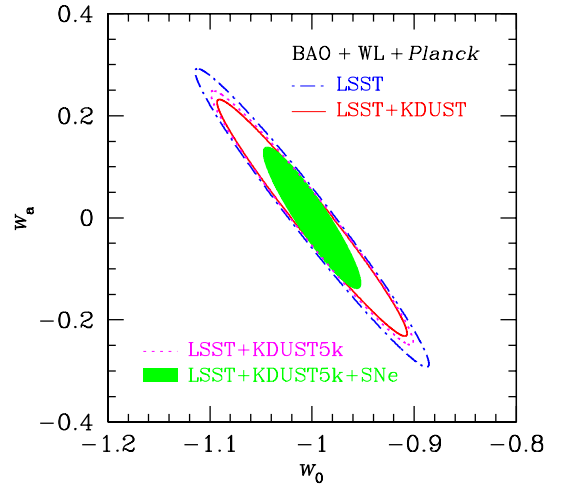}
\caption{Forecasts of 1$\sigma$ errors on the dark
energy EOS parameters $w_0$ and $w_{\rm a}$ for LSST using BAO+WL
(dash-dotted line), the combination of LSST and KDUST using BAO+WL
(solid line, see \autoref{tab:suv}), the combination of LSST and
half of KDUST (labeled as KDUST5k) using BAO+WL (dotted
line), and the combination of LSST and KDUST5k using BAO+WL+SNe
(shaded area). \label{fig:lstkd}}
\end{figure}

In summary, the parameter set includes 11 cosmological parameters
and 170 nuisance parameters. The cosmological parameters are
$w_0$, $w_{\rm a}$, the matter density $\omega_{\rm m}$, the baryon
density $\omega_{\rm b}$, the angular size of the sound horizon at
the last scattering surface $\theta_{\rm s}$, the curvature
parameter $\Omega_{\rm k}$, the scalar spectral index $n_{\rm s}$,
the running of the spectral index $\alpha_{\rm s}$, the primordial
Helium fraction $Y_{\rm p}$, the election optical depth $\tau$, and
the normalization of the primordial curvature power spectrum
$\Delta_{\rm R}^2$. Note that $Y_{\rm p}$ and $\tau$ are solely
constrained by the cosmic microwave background (CMB), which is
introduced as priors.
The nuisance parameters include 40 \phz{} bias parameters, 40 \phz{}
rms parameters, 40 galaxy clustering bias parameters, 30 galaxy
additive noise parameters, 10 shear additive noise parameters, and
10 parameters for shear calibration errors. We use multipoles
$40\le \ell \le 2000$ for WL and $40 \le \ell \le 3000$ for BAO.
In addition, we require $\Delta_\delta^2(\ell/D_A; z) < 0.4$ for BAO
to reduce the influence of nonlinear evolution. The lower cut in
$\ell$ is to minimize the dependence of the forecasts on particular
models of dark energy perturbation and the integrated Sachs--Wolfe
effect, which affect only very large scales.

\autoref{fig:kdall} presents the forecasts of 1$\sigma$ error
contours of the dark energy EOS parameters $w_0$ and $w_{\rm a}$ for
the ideal 10k survey WL (solid line), BAO (dotted line), SNe
(dot-dashed line),
joint BAO and WL (dashed line), and the three combined (shaded
area). Since different probes have different parameter degeneracy
directions in the full parameter space, including both cosmological
parameters and nuisance parameters (such as the \phz{} bias and rms
parameters), a joint analysis can reduce the error significantly.
One example is that the BAO technique can determine the curvature
parameter and the matter density far better than the SNe technique,
whereas the latter needs strong priors on $\Omega_{\rm K}$ and
$\omega_{\rm m}$ in order to place tight constraints on $w_0$ and
$w_{\rm a}$ \citep{linder05b,knox06c}. Another example is that the
WL technique is sensitive to the priors on the \phz{} parameters,
which can be calibrated by the BAO technique. By comparing the
results in \autoref{fig:kdall} with the LSST BAO+WL result in
\autoref{fig:lstkd}, one sees that the ideal 10,000 deg$^2$
\emph{ugrizyJH} survey can place comparable constraints on the
dark energy EOS parameters to the LSST survey.

We show the dark energy constraints from combinations of LSST with
the KDUST 10,000 deg$^2$ survey and LSST with half of the KDUST
survey (labeled as KDUST5k) in \autoref{fig:lstkd}. The KDUST5k
survey could be carried out by a 2.5m KDUST pathfinder in 12 years.
We assume that 5000 deg$^2$ high-resolution \emph{yJH}
imaging is enough to reach the systematic calibration floor, so the
improvement to the LSST survey outside the KDUST survey area is
kept the same for both proposals. Under this condition,
the WL+BAO constraints on $w_0$ and $w_a$ from LSST+KDUST5k are
nearly the same as those from LSST+KDUST; they improve the
LSST-alone dark energy task force \citep[DETF,][]{albrecht06b}
figure of merit (FOM) by 30\%. Adding SNAP-like SN data from
KDUST as well can significantly increase the DETF FOM; the result
is as good as LSST or LSST+KDUST joint BAO and WL constraints on
$w_0$ and $w_a$ in the absence of systematics.

The current constraint on $\{w_0,~w_a\}$ is \citep{zhao09},
\be\label{eq:w0wa_constraint}
w_0=-0.90^{+0.11+0.23}_{-0.11-0.22},~w_a=-0.24^{+0.56+0.98}_{-0.55-1.2}.
\ee derived from a joint Markov Chain Monte Carlo (MCMC) analysis of
the recently released SNe ``Constitution" sample
\citep{Hicken:2009dk}, the WMAP five-year
data~\citep{WMAP5}~\footnote{We have checked that the result is
largely unchanged if the WMAP 5 year data is replaced with WMAP 7
year data.}, and the Sloan Digital Sky Survey (SDSS) Luminous Red
Galaxy (LRG) sample \citep{Tegmark:2006az}. In this analysis, the
dark energy perturbation (DEP), which is important in the parameter
estimation~\citep{Zhao:2005vj,Fang:2008sn}, was consistently
included based on the treatment developed in \citep{Zhao:2005vj}.
The central values indicate that the `quintom'
scenario~\citep{quintom} is mildly favored, namely, the EOS today
$w(z)|_{z=0}=w_0>-1$, while EOS in the far past
$w(z)|_{z=\infty}=w_0+w_a<-1$. This is consistent with the recent
published result using the `Constitution' SNe
sample~\citep{cfa1,cfa2,cfa3,cfa4,cfa5}. The error bars in
\autoref{eq:w0wa_constraint} can be used to estimate the FOM of
current data, and we find that KDUST+LSST can improve the current
FOM by two orders of magnitude.

\subsection{Principal Component Analysis on $w(z)$}

To investigate the constraints on dark energy equation-of-state
$w(z)$ from the surveys in a model-independent way, we follow
\citep{Huterer:2002hy,Crittenden:2005wj} and employ a Principle
Component Analysis (PCA) approach.

We choose 40 uniform redshift bins, stretching to a maximum redshift
of $z=3$, and allow the high redshift ($z>3$) equation of state to
vary. To avoid $w(z)$ with infinite derivatives, each bin rises and
falls following a hyperbolic tangent function with a typical
transition width $dz$ of order 10\% of the width of a bin. We choose
a constant $w = -1.0$ as the fiducial model, which is consistent
with all the present data.
Since dark energy perturbations play a crucial role in the parameter
estimation, we use a modified
version of CAMB which allows us to calculate DEP for an arbitrary
$w(z)$ consistently~\citep{Zhao:2005vj}.

For the PCA, we calculate the Fisher matrices based on four kinds of
observables: supernovae, CMB
anisotropies, galaxy number counts (GC) correlation functions and WL
observations. We also include all the possible cross-correlations
among these, including CMB$\times$galaxy and CMB$\times$WL, which
are sensitive to the integrated Sachs-Wolfe effect, as well as
galaxy-weak lensing correlations.

We first calculate the Fisher matrices for each of the observables,
$F^{a}_{ij}$, where the indices $i,j$ run over the parameters of the
theory, in our case the binned $w_i(z)$.  We then find the
normalized eigenvectors and eigenvalues of this matrix
$\{e_i(z),\lambda_i\}$, and write
\begin{equation}
F = W^T \Lambda W,
\end{equation}
where the rows of $W$ are the eigenvectors and $\Lambda$ is a
diagonal matrix with elements $\lambda_i$. The Fisher matrix is an
estimate of the inverse covariance matrix we expect the data to give
us and the eigenvalues reflect how well the amplitude of each
eigenvector can be measured.  The true behavior of the equation of
state can be expanded in the eigenvectors as
\begin{equation}
w(z) = w_{\rm fid}(z) + \sum_{i=1}^N \alpha_i e_i (z)
\end{equation}
and the expected error in the recovered amplitudes is given by
$\sigma(\alpha_i) = \lambda_i^{-1/2}$.

For our forecasts, we assume three surveys: the ideal 10k survey,
LSST and KDUST+LSST as listed in \autoref{tab:suv}. We also combine
the Planck survey for CMB and SNAP SNe.
We assume a flat universe and marginalize over the intrinsic SN
magnitude $M$ and the galaxy bias parameters.

\begin{figure}[thp]
\centering
{\includegraphics[scale=0.75, ]{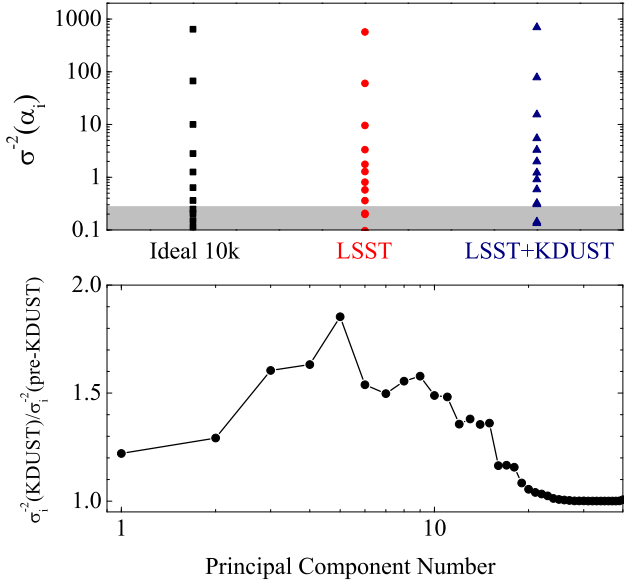}} \caption{Upper panel: The
eigenvalues ($\sigma^{-2}(\alpha_i)$) for the raw Fisher matrices
(no priors assumed) for ideal 10k survey, LSST, and LSST+KDUST.
The grey shaded
region shows the diagonal prior of $\sigma_{\rm m}\geq0.3$; Lower
panel: The ratio of the eigenvalues ($\sigma^{-2}(\alpha_i)$) for
LSST+KDUST to that for LSST.} \label{fig:eval}
\end{figure}

\begin{figure}[thp]
\centering
{\includegraphics[scale=1, ]{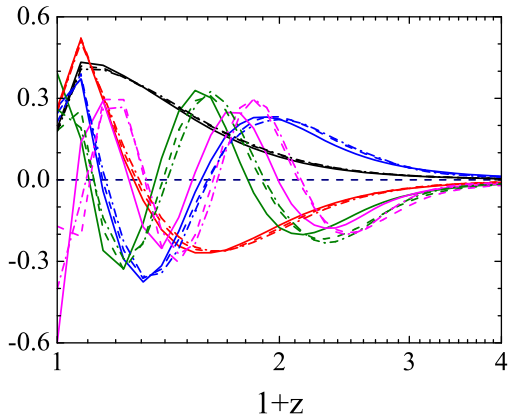}} \caption{The first five best
determined eigenvectors for the ideal 10k survey (dashed lines),
LSST (solid lines),
and LSST+KDUST (dash-dotted lines). No priors have been assumed and the
amplitudes are normalized to unity.
} \label{fig:evec}
\end{figure}

In the upper panel of \autoref{fig:eval}, we show the spectra of
eigenvalues $\sigma^{-2}(\alpha_i)$ of the Fisher matrices for these
three tomographic surveys, and in the lower panel, we show the ratio
of the eigenvalues for LSST+KDUST to that for LSST. As we can see,
the ideal 10k survey is competitive to LSST on the dark energy
constraints, and adding KDUST to LSST can improve the constraints on
the first few eigenmodes by as much as 85\%. The first five best
determined eigenvectors are shown in \autoref{fig:evec}. As we can
see, the $N$th eigenmode has $N-1$ nodes in $z$, and the `sweet
spot' -- the redshift where the uncertainty of $w(z)$ gets minimized
-- is at $z\sim0.2$. This is consistent with the analysis done in
\citep{Huterer:2002hy,Crittenden:2005wj}. We also find that adding
KDUST to LSST makes the eigenmodes stretch to slightly higher $z$
because of the higher redshift reach with the addition of KDUST
data.

As stated in \citep{Crittenden:2005wj}, all of the eigenvectors are
informative, no matter how large the error bars are, if we have no
prior knowledge of the possible $w(z)$ behaviors. However, even
without a physical model for $w(z)$, we would still be surprised if
$w(z)$ were much too positive ($w \gg 1/3$) or much too negative ($w
\ll -1$). This motivates us to choose some theoretical priors to
roughly separate the eigenmodes into those which are informative
relative to the priors and those that are not. To apply the
theoretical prior, we follow \citep{Crittenden:2005wj} to choose a
correlation function describing fluctuations of $w(z)$ away from the
fiducial model. This smoothness prior can filter out the high
frequency modes, while the low frequency modes remain unaffected.
Also, as long as there are sufficient bins compared to the
correlation length, the prior largely wipes out dependence on the
precise choice of binning.

As elaborated in \citep{Crittenden:2005wj}, the deviations of the
equation of state from its fiducial model can be encapsulated in a
correlation function:
\begin{equation}
 \xi_w (|z - z'|) \equiv \left\langle (w(z) - w_{\rm fid}(z))(w(z') - w_{\rm fid}(z')) \right\rangle.
\end{equation}
and the covariance matrix of the binned equation of state is then
\begin{eqnarray} \langle \delta w_i \delta w_j  \rangle =
\frac{1}{\Delta^2} \int_{z_i}^{z_i+\Delta} dz
\int_{z_j}^{z_j+\Delta}  dz' \, \xi_w (|z - z'|) .
\end{eqnarray} where the i$^{th}$ bin is from $z_i$ to $z_{i} + \Delta$,
and we assume that all bins have the same width $\Delta = z_{i+1} -
z_i$. The variance of the mean equation of state over all the bins
is, \begin{equation} \sigma_{\rm m}^2 =  \int_0^{z_{\rm max}} {dz}
\int_0^{z_{\rm max}} {dz' \over z_{\rm max}^2}\xi_w(z-z').
\end{equation} And we assume that $\xi_w
(z) =  \xi_w (0) /(1 + (z/z_c)^2).$ where $\xi_w (0)$ is the
variance of $w$ at any given point and $z_c$ is the correlation
length.

To apply the prior, we need to specify the correlation length $z_c$
and tune $\xi_w (0)$ so that the error in the mean, $\sigma_{\rm m}$
is in the range of $[0.2,0.5]$ in order to be consistent with the
observational uncertainty. We tried correlation lengths in the range
$0.1 \le z_c \le 0.4$, where the upper limit was beginning to be
strong enough to impact the observed modes for the SN. Then the
resulting prior takes the form,
\begin{equation}
{\cal{P}}_{\rm prior} \propto \exp{\left[-\frac{1}{2}({w}_i^{\rm
true} - {w}_i^{\rm fid})C_{ij}^{-1}({w}_j^{\rm true} - {w}_j^{\rm
fid})\right]}
\end{equation}
where $C_{ij} \equiv \langle \delta w_i \delta w_j \rangle$. This
prior naturally constrains the high frequency modes without over
constraining the lower frequency modes that are typically probed by
experiments.

Note that assuming no correlations among the bins is equivalent to
using a delta function for the correlation prior, e.g. $\xi(z) =
\xi_0 \delta(z).$ In such a case, one finds $\langle \delta w_i
\delta w_j \rangle = \xi_0 \delta_{ij}/\Delta $  and the mean
variance is $\sigma_{\rm m}^2 = \xi_0 /z_{\rm max}$.   Thus,
assuming a fixed total range, the bin variance should grow with the
number of bins $\langle \delta w_i^2 \rangle = \sigma_{\rm m}^2
N_{\rm bins}$ to keep the mean variance unchanged.

In order to evaluate the value added by a given survey, we use the
mean squared error (MSE) as a FOM. In the absence
of any priors, this is expected to be
\begin{equation}
{\rm MSE}  = {\rm Tr} \,F^{-1}.
\end{equation}
Taking into account priors, $F$ in this expression is replaced by
$C^{-1}+F$. The MSE arising from the prior alone is ${\rm MSE} =
{\rm Tr}~C$, and it is independent of the shape of the correlation
function. Adding more experimental data reduces the MSE, and the
amount by which the MSE is reduced can be seen as a measure of how
informative the experiment is.

In \autoref{tab:mse}, we vary $\xi_w(0)$ and $z_c$ while holding the
prior constraint on the mean fixed, $\sigma_{\rm m} = 0.3$. For the
case of a diagonal prior, the MSE for the prior alone is given by
$N_{\rm bins} \times \langle \delta w^2 \rangle = N_{\rm bins}^2
\sigma^2_{\rm m} = 144.$ Adding the data, one could significantly
improve the constraints on some of the eigenmodes, thus reduce the
MSE.  As the prior is diagonal, the new eigenmodes are also
eigenmodes of the prior with the same eigenvalue, given by  $N_{\rm
bins} \sigma^2_{\rm m} = 3.6$ (shaded region in the upper panels of
\autoref{fig:eval}). The reduction of the MSE thus roughly tells us
how many modes can be constrained compared to the prior. For
example, adding LSST reduces the MSE from 144 to 112, meaning that
there are $(144-112)/3.6 \simeq 9$ modes can be constrained by LSST,
and similarly, the ideal 10k survey is able to constrain 7 modes.
With KDUST+LSST, one can actually constrain 10 eigenmodes. These numbers
can also be counted in the upper panel of \autoref{fig:eval} above
the prior threshold shown in shade.

For a correlated prior, we can analogously put a lower bound on the
number of modes constrained by data using the MSE shown in
\autoref{tab:mse}. For instance, for the case of $z_c=0.1$ and
$z_c=0.4$, LSST combined with KDUST can at least constrain
$(35-18.1)/3.6\simeq4$ and $(11.5-3.7)/3.6\simeq2$ modes,
respectively. As we see, the number of new modes estimated in this
way is reduced as the prior correlation length is increased. This is
as expected -- as $z_c$ increases, more high frequency modes will be
filtered out by the smoothness prior.

\begin{table}
\centering
\caption{The mean squared error for various priors and experiments.
\label{tab:mse}}
\begin{tabular}{@{}l@{~~~}c@{~}c@{~}c@{~}c@{}c@{}}
\hline\hline
      & diagonal  & $\,\,z_c=0.1\,\,$  & $\,\,z_c = 0.4$ \,\,\\
\hline
no data                   & 144.0 & 35.0  & 11.5 \\
\hline
Ideal 10k                   & 116.0 [7] & 21.2  & 4.7 \\
LSST                        & 112.0 [9] & 19.9  & 4.5 \\
KDUST+LSST                    & 108.3 [10] & 18.1  & 3.7\\
\hline\hline
\end{tabular}
\tablecomments{The mean squared error is related to the number of
well constrained modes. The
priors are normalized so that $\sigma_{\rm m} =0.3$ and the
``diagonal'' prior means no correlations exist between bins.  For this
diagonal prior, we give in brackets the inferred number of modes
meaningfully constrained by the observations.}
\end{table}


\section{Conclusion} \label{sec:con}
Dome A offers a very competitive site for studying dark energy.
Given the amount of resources required to build a large telescope
and run a massive survey there, one must give the highest
priority to programs that cannot be easily carried out elsewhere. Thus,
a reasonable strategy is to focus on NIR imaging and collaborate
with other surveys for optical data. Using LSST as an example, we
show that a high-resolution 5000--10,000 deg$^2$ KDUST survey in
\emph{yJH} bands could improve LSST BAO+WL constraints on the dark
energy EOS parameters $w_0$ and $w_a$ by reducing the \phz{} and
shear measurement systematics. A SNAP-like SN sample plus a large
local and nearby SN sample from KDUST would
further boost the DETF FOM by more than a factor of two.

In addition to forecasts for the $w_0$--$w_a$ parametrization, we
also apply a PCA approach to investigate the constraints on the
dark energy EOS $w(z)$ in a model-independent way.
We find that regarding the number of the constrained eigenmodes
of $w(z)$, an ideal 10,000 deg$^2$ \emph{ugrizyJH}
survey, combined with \emph{Planck}, can constrain 7 eigenmodes,
while KDUST+LSST can allow us to constrain 3 more modes.

We have not discussed dark energy probes such as strong lensing,
cluster counting, and higher-order statistics of the same galaxy
and shear data, which could further tighten the constraints on
the dark energy EOS. Strong lensing constrains dark energy through
the time delay effect as well as counting of strong lenses. It is
also an excellent probe of dark matter halo structures and, hence,
can be used to measure dark matter particle properties.
With high-resolution imaging, one could
extract more cosmological information from strong lensing
observations. Therefore, Dome A could be particularly advantageous
for strong lensing studies.

Dark energy forecasts depend crucially on the assumed properties of
the survey data, including all the systematics. Dome A has many
advantages over other ground sites and has an environment close to
that in space. Hence, we use well-studied LSST and SNAP as
references to make crude estimates of the data for this investigation.
Further work and detailed modeling are needed to give a more
realistic assessment of the Dome A site for studying dark energy.

\acknowledgments We thank David Bacon, Rob Crittenden, Kazuya Koyama, Bob Nichol and Levon Pogosian for useful
discussions. GZ is supported by the ERC grant. HZ is supported by
the Bairen program from the Chinese Academy of Sciences and
the National Basic Research Program of China grant No. 2010CB833000. The work of L. Wang is partially supported
by NSF grant AST-0708873. XZ is supported
in part by the NSF of China.



\end{document}